\documentstyle[12pt]{article}
\def\a{\alpha}
\def\b{\beta}
\def\g{\gamma}
\def\d{\delta}
\def\e{\epsilon}
\def\et{\eta}

\def\k{\kappa}
\def\l{\lambda}
\def\m{\mu}
\def\n{\nu}
\def\t{\tau}
\def\G{\Gamma}

\def\L{\Lambda}
\def\s{\sigma}
\def\th{\theta}
\def\bra{\langle}
\def\ket{\rangle}
\def\eq{eq.~}
\def\MPL #1 #2 #3 {{\em Mod.~Phys.~Lett.}~{\bf #1}\ #2 (#3)}
\def\NPB #1 #2 #3 {{\em Nucl.~Phys.}~{\bf#1}\ (#2) #3}
\def\PLB #1 #2 #3 {{\em Phys.~Lett.}~{\bf B #1}\ (#2) #3}
\def\PR #1 #2 #3 {{\em Phys.~Rep.}~{\bf #1}\ (#2) #3}
\def\PRD #1 #2 #3 {{\em Phys.~Rev.}~{\bf D #1}\ (#2) #3}
\def\PRL #1 #2 #3 {{\em Phys.~Rev.~Lett.}~{\bf #1}\ (#2) #3}
\def\RMP #1 #2 #3 {{\em Rev.~Mod.~Phys.}~{\bf #1}\ (#2) #3}
\def\ZP #1 #2 #3 {{\em Z.~Phys}.~{\bf#1}\ (#2) #3}
\def\PTP #1 #2 #3 {{\em Prog.~Theor.~Phys}.~{\bf#1} \ (#2) #3}

\def\Vcb{V_{cb}}

\def\GSW_sign{}

\def\mv{m_{K^\ast}}
\def\MP{m_K}
\def\to{\rightarrow}

\def\p{\prime}
\def\shat{{\hat s}}

\def\BXll{B\to X_s\ell^+\ell^-}
\def\BKll{B\to K\ell^+\ell^-}
\def\BKsll{B \to K^\ast \ell^+\ell^-}
\def\dd{{\cal D}}
\def\Ce{C_8^{eff}}
\def\Cn{C_9}
\def\Cs{C_7}

\def\Cep{C_{8^\p}}
\def\Cnp{C_{9^\p}}
\def\Csp{C_{7^\p}}
\def\fp{F_+}
\def\fm{F_-}
\def\ft{F_T}
\makeatletter
\newcommand{\nn}{\nonumber}
\newcommand{\bd}{\begin{document}}
\newcommand{\ed}{\end{document}}
\newcommand{\bc}{\begin{center}}
\newcommand{\ec}{\end{center}}
\newcommand{\be}{\begin{eqnarray}}
\newcommand{\ee}{\end{eqnarray}}
\newcommand{\eqn}{\global\def\theequation}
\newcommand{\sw}{sin^2 \theta_W}
\newcommand{\fbd}{f_B}
\renewcommand{\thefootnote}{\alph{footnote}}
\def\figcap{\section*{Figure Captions\markboth
     {FIGURECAPTIONS}{FIGURECAPTIONS}}\list
     {Figure \arabic{enumi}:\hfill}{\settowidth\labelwidth{Figure 999:}
     \leftmargin\labelwidth
     \advance\leftmargin\labelsep\usecounter{enumi}}}
\let\endfigcap\endlist \relax
\def\reflist{\section*{References\markboth  
     {REFLIST}{REFLIST}}\list
     {[\arabic{enumi}]\hfill}{\settowidth\labelwidth{[999]}
     \leftmargin\labelwidth
     \advance\leftmargin\labelsep\usecounter{enumi}}}
\let\endreflist\endlist \relax
\renewcommand{\thefootnote}{\alph{footnote}}
\begin{document}
\tolerance=10000
\begin{titlepage}
\begin{flushright}
{\normalsize     NHCU-HEP-96-14
\\ hep-ph/9608466}
%\\ July~1996}    
\end{flushright}
 \null
 \vskip 0.5in   
\begin{center}  
 \vspace{.15in}  
{\Large {\bf
 Lepton Polarization Asymmetry in $B\to K^{(*)} l^+l^-$
}}\\
  \par
 \vskip 2.5em   
 {\large
  \begin{tabular}[t]{c}
        {\bf C.~Q.~Geng and C.~P.~Kao}
\\
\\
   {\em   Department of Physics, National Tsing Hua University} \\
  {\em  Hsinchu, Taiwan, Republic of China}
   \end{tabular}}
 \par \vskip 5.0em
 {\large\bf Abstract}
\end{center}
\setlength{\baselineskip}{5ex}

We investigate the longitudinal lepton polarization asymmetry 
in the exclusive  processes  $\BKll$ and $\BKsll$.  
We include both short and long-distance contributions to the 
asymmetry in our discussions. We find that average values of the 
polarization asymmetries of the muon and tau 
for $B\to K^{(*)}\m^+\m^-$ and $B\to K^{(*)}\t^+\t^-$
are $-0.8\ (-0.7)$ and $-0.2\ (-0.5)$,
respectively.

\end{titlepage}

\newpage
\setlength{\baselineskip}{5ex}

\section{Introduction}

~~~As is well known, the study of B-meson physics is important
in the determination of the elements of the Cabibbo-Kobayashi-Maskawa
(CKM) \cite{CKM} matrix of charged-current weak couplings. In addition,
it could shed light on physics beyond the standard model.
Recently, much interest \cite{Hewett2} has been centered on rare B-meson 
decays induced by the flavor changing neutral current $b\to s$ 
transition due to the CLEO measurement of the radiative $b\to s\gamma$ 
decay \cite{CLEO}.
In the standard model, these rare processes do not occur 
at tree level and appear only at the quantum (loop) level.
The short distance (SD) contributions to the decays involving the $b\to s$ 
transition are dominated by loops with the top quark and 
they are basically free of the uncertainties in the CKM parameters.
The rare B-meson decays are thus a good probe of heavy top quark physics
as well as physics beyond the standard model.

It has been pointed out by Ali $et\ al.$ \cite{ali}
that, in the standard model, the measurement of the forward-backward 
asymmetry of the dileptons in the inclusive decays $b\to sl^+l^-$ 
provides information on the short-distance contributions dominated by 
the top quark loops. Recently, it has been emphasized by Hewett \cite{JLH}
that the longitudinal lepton polarizations, which are another parity
violating observables, are also important asymmetries. In particular,
the tau polarization in $b\to s\tau^-\tau^-$ mode could be accessible
to the B-Factories currently under construction.
It is interesting to note that the dilepton forward-backward asymmetry
of the exclusive decays $B\to M l^+l^-$ is identically zero when $M$ are
pseudoscalar mesons such as $\pi$ and $K$ but nonzero for $M$ being
vector mesons such as $\rho$ and $K^*$.
However, the longitudinal lepton polarizations for the exclusive modes
are nonzero for both cases of pseudoscalar and vector mesons.
In this paper we will examine these lepton polarization asymmetries 
for the exclusive decays of $B\to Kl^+l^-$ and $B\to K^*l^+l^-$,
in which we only concentrate on the muon and tau dileptonic modes since 
the electron polarization is hard to be measured experimentally.
In our discussions, We will use the results of the relativistic quark 
model by the light-front formalism \cite{Jaus,JAW} on 
the form factors in the hadronic matrix elements between the B-meson and
the kaons.

The paper is organized as follows. In Sec. 2, we
give the general effective Hamiltonian in the standard model for the rare 
dilepton decays of interest. 
We study the exclusive decays of $\BKll$ and $\BKsll$ in Sec. 3 and
Sec. 4, respectively.
Our conclusions are summarized in Sec. 5.

\section{Effective Hamiltonian}

~~~The effective  Hamiltonian  relevant to $b\to s l^+ l^-$ 
based on an operator product expansion is given by \cite{GW}
\be
\label{heff}
{\cal H}_{eff} = \frac{4 G_F \l_t}{\sqrt{2}} \,\sum_{i=1}^{10}
C_i(\m) \, O_i(\m)\,,
\ee
where $G_F$ denotes the Fermi constant, $\l_t=V_{tb}V^*_{ts}$ are
the products of the CKM matrix elements. 
  The $O_i(\m)$ are the following operators 
\be
\label{basis}
O_1 &=& \left( \bar{s}_\a \g^\m L b_\a \right) \
        \left( \bar{c}_\b \g_\m L c_\b \right) \;, \nonumber \\
O_2 &=& \left( \bar{s}_\a \g^\m L b_\b \right) \
        \left( \bar{c}_\b \g_\m L c_\a \right) \; , \nonumber \\
O_3 &=& \left( \bar{s}_\a \g^\m L b_\a \right) \
        \left[ (\bar{u}_\b \g_\m L u_\b)+\cdots+  
               (\bar{b}_\b \g_\m L b_\b)\right] \;, \nonumber \\
O_4 &=& \left( \bar{s}_\a \g^\m L b_\b \right) \
        \left[ (\bar{u}_\b \g_\m L u_\a)+\cdots+  
               (\bar{b}_\b \g_\m L b_\a)\right] \;, \nonumber \\
O_5 &=& \left( \bar{s}_\a \g^\m L b_\a \right) \
        \left[ (\bar{u}_\b \g_\m R u_\b)+\cdots+  
               (\bar{b}_\b \g_\m R b_\b)\right] \;, \nonumber \\
O_6 &=& \left( \bar{s}_\a \g^\m L b_\b \right) \
        \left[ (\bar{u}_\b \g_\m R u_\a)+\cdots+  
               (\bar{b}_\b \g_\m R b_\a)\right] \;, \nonumber \\
O_7 &=& \frac{e}{16 \pi^2} \, m_b \, (\bar{s}_\a \s^{\m \n} R b_\a)
         \, F_{\m \n} \;, \nonumber \\
O_8 &=& \frac{e^2}{16 \pi^2} \, (\bar{s}_\a \g^\m L b_\a) \bar{l}
         \g_\m l \;,  \nonumber \\
O_9 &=& \frac{e^2}{16 \pi^2} \, (\bar{s}_\a \g^\m L b_\a)
         \bar{l} \g_\m \g_5 l \; , \nonumber \\
O_{10} &=& \frac{g}{16 \pi^2} \, m_b \, (\bar{s}_\a \s^{\m \n} T^a_{\a\b}
          R b_\b)
         \, G^a_{\m \n} \;,
\ee
where $R(L)=(1\pm\g_5)/2$.
Here $F_{\m\n}$ and $G_{\m\n}^a$ are the electromagnetic and strong
interaction field strength tensors, and $e$ and $g$ are 
the corresponding coupling constants, respectively. 
In the standard model with the dimension six operator basis, 
$O_1$ and $O_2$ are current-current operators, $O_3,\cdots, O_6$ are 
usually QCD penguin operators, $O_7$ and $O_{10}$ are magnetic penguin 
operators, $O_8$ and $O_9$ are semileptonic electroweak operators.
The $C_i$ in \eq(\ref{heff}) are the 
corresponding Wilson coefficients.
They are calculated first at a renormalization scale $M_W$ and
then scaled down to a mass of order $m_b$ using the
renormalization group equations.
It is known that the coefficients $C_3,\cdots,C_6$ are small and thus the 
contributions of the corresponding operators can be neglected.
Moreover the operator $O_{10}$ does not contribute to the decay of
$b\to s l^+l^-$. Hence the relevant operators of our study are 
$O_1$,$O_2$,$O_7$,$O_8$ and $O_9$, respectively.
At the scale $\m=M_W$, one has \cite{Inami}
\be
\nn
C_1(M_W)&=&0, 
\nn\\
C_2(M_W)&=&-1, 
\nn\\
C_7(M_W)&=&\frac{1}{2} A(x) ,
\nn\\
C_8(M_W)&=&\frac{1}{\sin^2 \th_W} B(x)+\frac{-1+4\sin^2\th_W}
{\sin^2 \th_W}C(x)+D(x)-\frac{4}{9}, 
\nn\\
C_9(M_W)&=&\frac{-1}{\sin^2 \th_W}B(x)+\frac{1}{\sin^2\th_W}C(x).
\ee
where $x=m_t^2/M_W^2$ and 
\be
A(x)&=&x\left[\frac{2x^2/3+5x/12-7/12}{(x-1)^3}-
       \frac{3x^2/2-x}{(x-1)^4}\ln x \right],
\nn\\
B(x)&=&\frac{1}{4}\left[\frac{-x}{x-1}+\frac{x}{(x-1)^2}\ln x\right],
\nn\\
C(x)&=&\frac{x}{4}\left[\frac{x/2-3}{x-1}+\frac{3x/2+1}{(x-1)^2}
       \ln x \right], 
\nn\\
D(x)&=&\frac{-19x^3/36+25x^2/36}{(x-1)^3}+
       \frac{-x^4/6+5x^3/3-3x^2+16x/9-4/9}{(x-1)^4}\ln x .\
\ee
Because $O_1$ and $O_2$ produce dilepton via virtual (vector)
photon, they can be incorporated into $O_8$ in the later calculation.
  From \eq(\ref{heff})  and \eq(\ref{basis}), we obtain the amplitude for
the inclusive process $\BXll$ \cite{JLH},
\be
\label{amp}
M  =  { G_F\a\over\sqrt 2\pi} \l_t\left( \Ce 
\bar s_L\g_\m b_L {\bar l}\g^\m l+\Cn\bar s_L\g_\m
b_L {\bar l}\g^\mu\g_5 l
  -2\Cs m_b\bar s_L i\s_{\m\n}{q^\n\over q^2}b_R {\bar l}\g^
\m l  \right).
\ee
The coefficients $\Ce,\Cn$ and $\Cs$ are given by \cite{Greub}
\be
\label{c8eff}
C_7(m_b)&=&\et^{-16/23}
\left\{C_7(M_W)-\frac{58}{135}(\et^{10/23}-1)C_2(M_W)
\right. 
\nonumber \\ 
&&
\left.  -\frac{29}{189}(\et^{28/23}-1)C_2(M_W) \right \}\,,
\nn\\
\Ce(m_b)&=&C_8(m_b)+[3C_1(m_b)+C_2(m_b)]\nonumber\\ 
&&\hspace{1.6cm} \times\left(h(\hat{m}_c,\shat) 
 -\frac{3}{\a^2} \k
\sum_{V_i=J/\psi,\psi^\p} \frac{\pi\G(V_i \to ll)M_{V_i}}{
q^2-M_{V_i}+iM_{V_i}\G_{V_i}}\right)\,,
\nn\\
C_9(m_b)&=&C_9(M_W)\,,
\ee
where
\be
\nn
C_1(m_b)&=&\frac{1}{2}(\et^{-6/23}-\et^{12/23})C_2(M_W),\\
\nn
C_2(m_b)&=&\frac{1}{2}(\et^{-6/23}+\et^{12/23})C_2(M_W), \\
C_8(m_b)&=&C_8(M_W)+\frac{4\pi}{\a_s(M_W)}\left\{-\frac{4}{
33}\left[1-\et^{-11/23}\right] \right.\nonumber\\&&\hspace{4cm} 
\left. +\frac{8}{87}\left[
1-\et^{-29/23}\right]\right\}C_2(M_W)\,.
\ee
Here
$\hat{m}_c=m_c/m_b$,\,$\et=\a_s(m_b)/\a_s(M_W)$,
$\a=e^2/4\pi$, $\shat=q^2/m^2_b$ with $q^2$ being the 
invariant mass of the dilepton, and 
$h(\hat{m}_c,\shat)$ arising from the one-loop contributions  
of $O_1$ and $O_2$ is given by 
\be
h(z,\shat)=-\frac{4}{9}\ln z^2+\frac{8}{27}+\frac{4}{9}y-
\frac{2}{9}(2+y)\sqrt{|1-y|}
\ \ \ \ \ \ \ \ \ \ \ \ \ \
\ \ \ \ \ \ \ \ \ \ \ \ \ \
\ \ \ \ \ \ \ 
\nonumber \\
\ \ \ \ \ \ \ \ \ \ \
\times \left\{\Theta (1-y)
 \left[ \ln\frac{1+\sqrt{1-y}}{1-\sqrt{1-y}}
+i\pi\right] +
\Theta (y-1)2\,arctan\left(\frac{1}{\sqrt{y-1}}\right) \right\},\
\ee
where $y\equiv 4z^2/\shat$.
The last term of $\Ce(m_b)$ in \eq (\ref{c8eff}) is the long-distance (LD) 
contribution mainly
due to the $J/\psi$ and $\psi^\p$ resonances \cite{LGJ}, and 
the factor $\k$ 
must be chosen such that $\k(3C_1(m_b)+C_2(m_b))\simeq -1$
in order to correctly reproduce the branching ratio \cite{ali}
\be
BR(B\to J/\psi X\to Xl\bar{l})=BR(B\to J/\psi X)BR(J/\psi \to l\bar{l}).
\ee

\section{Longitudinal lepton polarization in  $\BKll$}
  
~~~The hadronic matrix elements of the operators $O_1$, $O_2$, $O_7$, 
$O_8$ and $O_9$ between the B-meson and the pseudoscalar K-meson are given 
in terms of form factors as follows \cite{Jaus,JAW,Greub,FF} 
\be
\label{kv}
 && \bra p_K | \bar{s}
\g_\m (1 \mp \g_5) b|p_B \ket =
F_+(q^2)P_{\m}+F_-(q^2)q_{\m},
\ee
\be
\label{kt}
&& \bra p_K | \bar{s}
i \s_{\m \n} q^\n (1 \pm \g_5) b |p_B \ket =
\frac{1}{m_B+m_K}[P_{\m}q^2-(m_B^2-m_K^2)q_{\m}]F_T(q^2),
\ee
where $P\equiv p_K+p_B$, and $q\equiv p_K-p_B$.
In this paper, we use the form factors given by the relativistic constituent 
quark model in refs. \cite{Jaus} and \cite{JAW}.
 The model is based on the light front formalism,
in which the form factors $\fp$,\,$\fm$ and $\ft$ are taken to be 
approximately as \be
F(q^2)=\frac{F(0)}{1-q^2/\L^2_1+q^4/\L^4_2}.
\ee
Here the parameters $\L_1$ and $\L_2$ are determined by the 
first and second derivative of $F(q^2)$ at $q^2=0$.
The values of $F(0)$ and $\L_1,\L_2$ for the various form factors used 
can be found in Table I of ref. \cite{JAW}.
When lepton mass (for $l=e$ or $\m$) is negligible,
 $q_\m$ terms in eqs. (\ref{kv}) and (\ref{kt})
give no contributions to the decay rate and which is the case studied in 
refs. \cite{Jaus} and \cite{JAW}. 
However, the contributions have to be considered for the tau dileptonic
channel of $B\to K \tau^+ \tau^-$. In this case, one cannot take $\fm$ as 
zero.  The form factor $\fm$ has been examined in ref. \cite{hwcw}, and it
is found to be 
\be
\label{efm1}
\fm&\simeq&-\fp(m_B^2+\MP^2)/(m_B^2-\MP^2)
\,.
\ee
$\fm$ can be extracted from other methods,
 such as that 
from Heavy Quark Symmetry (HQS) \cite{Isgur}. In this method, one has
\be
\label{efm2}
\fm&=&\fp+\frac{2m_B \ft}{m_B+\MP}\,.
\ee

The differential decay rate is given by
\be
\label{dr:kll}
\frac{d\G(\BKll)}{d\shat}&=&\frac{G_F^2 |\l_t|^2 m_B^5 \a^2}{3 \!\cdot
 2^9 \pi^5}
\dd\phi^{1/2}
\left[ [|\Ce \fp - \frac{2 \Cs \ft}{1+\sqrt{r}}|^2\right.
\nn\\
&&+|\Cn \fp|^2]\phi\left(1+\frac{2t}{\shat}\right)
+12|\Cn|^2 t[(1+r-\shat/2)\fp^2
\nn\\
&&\left.+(1-r)\fp\fm+\frac{1}{2}\shat\fm^2]\right],
\ee
where 
\be 
\label{e14}
&&t=m_l^2/m_B^2,\quad r=\MP^2/m_B^2,\quad \shat=q^2/m_B^2,\quad 
\nonumber\\&&
\phi=(1-r)^2-2\shat(1+r)+\shat^2,\quad 
\dd=\sqrt{1-4t/\shat}.
\ee
We note that the rate in \eq (\ref{dr:kll}) is a general form with 
$m_l\neq 0$ and it agrees with that given in ref. \cite{Du}.
The differential branching ratios, 
$dB/d\shat\equiv d\G/(\G_{tot}d\shat)$, 
as a function of $\shat$ for $B\to K\m^+\m^-$ and 
$B\to K\t^+\t^-$ decays are displayed in Fig. 1a and Fig. 1b, respectively.
Here we have used that $m_t=180\ GeV$ and $\G_{tot}\simeq m_b^5 G_F^2 
|\Vcb|^2/(64\pi^3)$ with $m_B\simeq m_b=5\ GeV$.
The dashed and solid lines in Figs. 1a and 1b represent the results
with and without the LD contributions
from the resonance states, respectively.
It is noted that the solid line for the decay of $B\to K\t^+\t^-$
in Fig. 1b is similar to the corresponding figure in ref. \cite{Du}.

For the dilepton decays of the B-meson,
the longitudinal polarization asymmetry (LPA) of the lepton, $P_L$, is 
defined as 
\be
\label{ePA}
P_L(\shat)=\frac{d\G_{h=-1}/d\shat-d\G_{h=1}/d\shat}{
d\G_{h=-1}/d\shat+d\G_{h=1}/d\shat}
\ee
where $h=+1(-1)$ means right (left) handed $\ell^-$
in the final state
and $d\G/d\shat$ means the differential
 decay rate of the B meson. 
  In the standard model, the polarization asymmetries in 
\eq (\ref{ePA}) for $\BKll$ and 
$\BKsll$ come from the interference of the vector or 
magnetic moment and axial-vector operators.
For the decay of $\BKll$, it is found that
\be
\label{LPA1}
 P_L(\shat)=2\dd\phi \fp \Cn
(\fp Re\Ce-2 \frac{\ft}{1+\sqrt{r}} \Cs )/R,
\ee
where $R$ is given by
\be
\label{dr:K}
R&=&[|\Ce \fp - \frac{2 \Cs \ft}{1+\sqrt{r}}|^2
+|\Cn \fp|^2]\phi\left(1+\frac{2t}{\shat}\right)
+12|\Cn|^2 t[(1+r-\shat/2)\fp^2
\nonumber\\&&
+(1-r)\fp\fm+\frac{1}{2}\shat\fm^2]\,.
\ee

In Figs. 2a and 2b, taking $m_t=180\ GeV$,
we show the LPAs for $\BKll$ as a function of $\shat$ 
with $l=\mu$ and $\tau$, respectively. 
The LPAs for both cases vanish at the thresholds and oscillate in  the
resonance regions of $q^2\simeq M^2_{\psi(\psi^\p)}$ and they also reach zero
at the end points.
However, we note that if $m_l=0$, the LPA in \eq(\ref{LPA1}) 
has a finite value at the maximal end point of $\shat_{max}=(1-m_K/m_B)^2$.
 From Figs. 2a and 2b, we find that, away from the resonance regions and end 
points, $P_L(\shat)$ of the muon dilepton channel is around $-0.9$, while
that of the tau is between $-0.2$ and $-0.3$ for $0.6\leq\shat\leq 0.8$.
The average values of the LPAs are $-0.8$ and $-0.2$, respectively.
Finally, we remarks that Figs. 1 and 2 were drew by using the 
form factor $\fm$ in \eq (\ref{efm1}). The curves in Figs. 1 
and 2 are nearly unchanged if one uses the form factor of $\fm$ 
formulated in \eq (\ref{efm2}).

\section{Longitudinal lepton polarization in  $\BKsll$}

~~~The hadronic matrix elements of the operators in 
\eq (\ref{amp}) between the external states $B$ and $K^*$ for
the exclusive decay of $B\to K^\ast l^+l^-$ are \cite{JAW}, 
\be
\label{eq218}
 && \bra p_{K^\ast} | \bar{s}
\g_\m (1 \mp \g_5) b|p_B \ket =
\frac{1}{m_B+m_{K^\ast}} \,
\left[ i V(q^2)
 \epsilon_{\m \n \a \b} \e^{\ast \n} P^\a
q^\b \right. \nonumber \\
&& \left.
\pm A_0(q^2) (m_B^2-m_{K^\ast}^2) \e^\ast_\m \pm A_+(q^2)
 (\e^\ast P) P_\m \pm
A_-(q^2) (\e^\ast P) q_\m \right]\,,
\ee
\be
\label{eq219}
\bra p_{K^\ast} | \bar{s}
i \s_{\m \n} q^\n (1 \pm \g_5) b |p_B \ket =
 i g(q^2)
\epsilon_{\m \n \a \b} \e^{\ast \n} P^\a q^\b \pm 
\ \ \ \ \ \ \ \ \ \ \ \ \ \ \ \ \ \ \ \ \ \ \ \
\nonumber \\
\ \ \ \ \ \ \ a_0(q^2) \,
(m_B^2 - m_K^2) \left[ \e^\ast_\m - \frac{1}{q^2} (\e^\ast q) q_\m
\right] \pm
a_+(q^2) \,
(\e^\ast P) \left[ P_\m - \frac{1}{q^2} (Pq) q_\m \right]\,,
\ee
where $P=p_B + p_{K^\ast}$ and
$q=p_B - p_{K^\ast}$.
The terms corresponding to $A_-$ in eqs. (\ref{eq218}) and (\ref{eq219})
are important only for the mode of $B\to K^*\tau^+ \tau^-$.
Similar to \eq (\ref{efm1}), one has that \cite{hwcw} 
\be
\label{eAm}
A_-&\simeq& -A_+(m_B^2+\mv^2)/(m_B^2-\mv^2)
\,.
\ee

 From eqs. (\ref{amp}), (\ref{eq218}) and (\ref{eq219}), 
 we  obtain the differential decay rate of $\BKsll$ as
\be
\label{dr:ksll}
  \frac{d\G(\BKsll)}{d\shat}=\frac{G_F^2                           m_B^5
|\l_t|^2}{8\pi^3}\frac{\a^2}{16\pi^2} \dd   \phi^{1/2}      \left[
(1+\frac{2t}{\shat})  \left( \frac{\shat}{m_B^2}  \a + \frac{\b}{3} \phi
\right) + t\d \right], 
\ee
where the definitions of $\a$ and $\b$ are the same as in ref. \cite{Greub}
with choosing zero values of $\Csp,\Cep$ and $\Cnp$, and
\be
\d &=&\frac{|\Cn|^2}{2(1+\sqrt{r})^2}\left\{ -2\phi|V|^2 - 3(1-r)^2|A_0|^2
+ \frac{\phi}{4r} \left[2(1+r)-\shat\right]|A_+|^2 \right. \nonumber \\
   && \left.+\frac{\phi\shat}{4r}|A_-|^2+\frac{\phi}{2r}(1-r)Re(
      A_0 A_+^\ast + A_0 A_-^\ast + A_+ A_-^\ast) \right\}. 
\ee
The forms of $r$ and $\phi$ are the same as that in \eq (\ref{e14})
with the replacement $\MP\to\mv$. 
The differential branching ratios
as a function of $\shat$ for $B\to K^*\m^+\m^-$ and 
$B\to K^*\t^+\t^-$ decays are shown in Fig. 3a and Fig. 3b, 
respectively, with $m_t=180\ GeV$.
In \eq (\ref{dr:ksll}), the lepton mass is kept to be nonzero
and it is consistent with the result given in
\cite{Greub} for the $\BKsll$ decay by taking the limit of $m_l=0$ .

 From \eq (\ref{ePA}), the LPA in $\BKsll$ is found to be
\be
\label{e25}
P_L(\shat)
&=&\frac{\dd}{3}
\frac{\Cn}{1+\sqrt{r}}
\left({Re\Ce\over 1+\sqrt{r}}S_{av}-{2\Cs\over\shat}S_{at}
\right)/R, 
\ee
where
\be
R&=&(1+\frac{2t}{\shat})\left( \frac{\shat}{m_B^2} \a + \frac{\b}{3} \phi
\right) + t\d\,, \nn\\
S_{av}&=&V^2 F_V+A_0^2 F_A^0+A_+^2 F_A^+ + A_0 A_+ F_A^{0+}\,,
\nonumber\\
S_{at}&=&gV F_V+a_0 A_0 F_A^0+{a_0 A_+ + a_+ A_0\over 2} F_A^{0+}+
       a_+ A_+ F_A^+\,,
\ee
with
\be
F_V&=&\shat\phi,\nonumber\\
F_A^0&=&\frac{1}{8 r} (1-r)^2 ((1-r)^2-2\shat+10
     r \shat+\shat^2), \nonumber\\
F_A^+&=&\frac{1}{8 r}\phi^2,\nonumber\\
F_A^{0+}&=&\frac{1}{4 r}\phi (1-r)
           (1-r-\shat).
\ee 
In Figs. 4a and 4b, we plot the LPAs as a function of $\shat$ 
for $B\to K^*\m^+\m^-$ and $K^*\t^+\t^-$, respectively, with $m_t=180\ GeV$. 
 From Fig. 4, we see that $P_L(B\to K^*l^+l^-)$ for both $l=\m$ 
and $\t$  vanish at the minimal end point of $\shat=4m_l^2/m_B$ due to 
the kinematic factor $\dd$ in \eq (\ref{e25}).
However, in contrast with the cases in $\BKll$ shown in Figs. 2a and 2b, 
$P_L(\BKsll)$ with $l=\m$ and $\t$ are not zero at the maximum of 
$\shat_{max}=(1-m_{K^*}/m_B)^2$.
This is due to the fact that the numerator in \eq({\ref{LPA1}) has
extra zero from $\phi$ at $\shat_{max}$ whereas it does not contain 
$\phi$ in \eq (\ref{e25}).
Similar to the cases of $\BKll$, $P_L(B\to K^*\m^+\m^-)$ 
has a large negative value over most of the allowed range of $\shat$,
with an average value $<P_L(B\to K^*\m^+\m^-)>=-0.7$
while for the $\t^+\t^-$ channel, the average tau polarization is
$<P_L(B\to K^*\t^+\t^-)>=-0.5$.
Finally, we note that our results in Figs. 3 and 4 are insensitive
to the values of $A_-$ predicted in the different form factor models.

\section{Conclusions}

~~~We have investigated the longitudinal polarization 
asymmetries of the muon and tau in the exclusive processes of $\BKll$ 
and $\BKsll$. 
The average values of the polarization asymmetries of the muon and tau 
are found to be $-0.8$ and $-0.2$ for $B\to K\m^+\m^-$ and $B\to K\t^+\t^-$,
and $-0.7$ and $-0.5$ for $B\to K^*\m^+\m^-$ and $B\to K^*\t^+\t^-$,
respectively. These polarization asymmetries provide valuable information on
the flavor changing loop effects in the standard model.
 From Figs. 1 and 3, we find that the total integrated branching ratios of
$B\to K^{(*)}l^+l^-$ are $0.5\ (1.4)\times 10^{-6}$ and
$1.3\ (2.2)\times 10^{-7}$ 
with $l=\m$ and $\t$, respectively. Experimentally, to measure an 
asymmetry $A$ of a decay with the branching ratio $B$ at the $n\sigma$ 
level, the required number of events is $N=n^2/(BA^2)$.
For example, to observe the tau polarizations at the both 
exclusive channels of $B\to K^{(*)}\t^+\t^-$, one need at least
$1.8\times 10^{7}n^2$ $B\bar{B}$ decays.
Therefore, at the B-Factories under construction, some of the 
asymmetries could be accessible. 

%\vspace{3cm}

\begin{center}
{\large\bf Acknowledgments}
\end{center}
We thank C.W. Hwang for showing their recent results \cite{hwcw}
on the hadronic form factors to us.
This work is supported in part by the National Science Council of
the Republic of China under grant NSC-85-2112-M-007-031.

\newpage
\setlength{\baselineskip}{3ex}

\begin{figure}[h]
\includegraphics{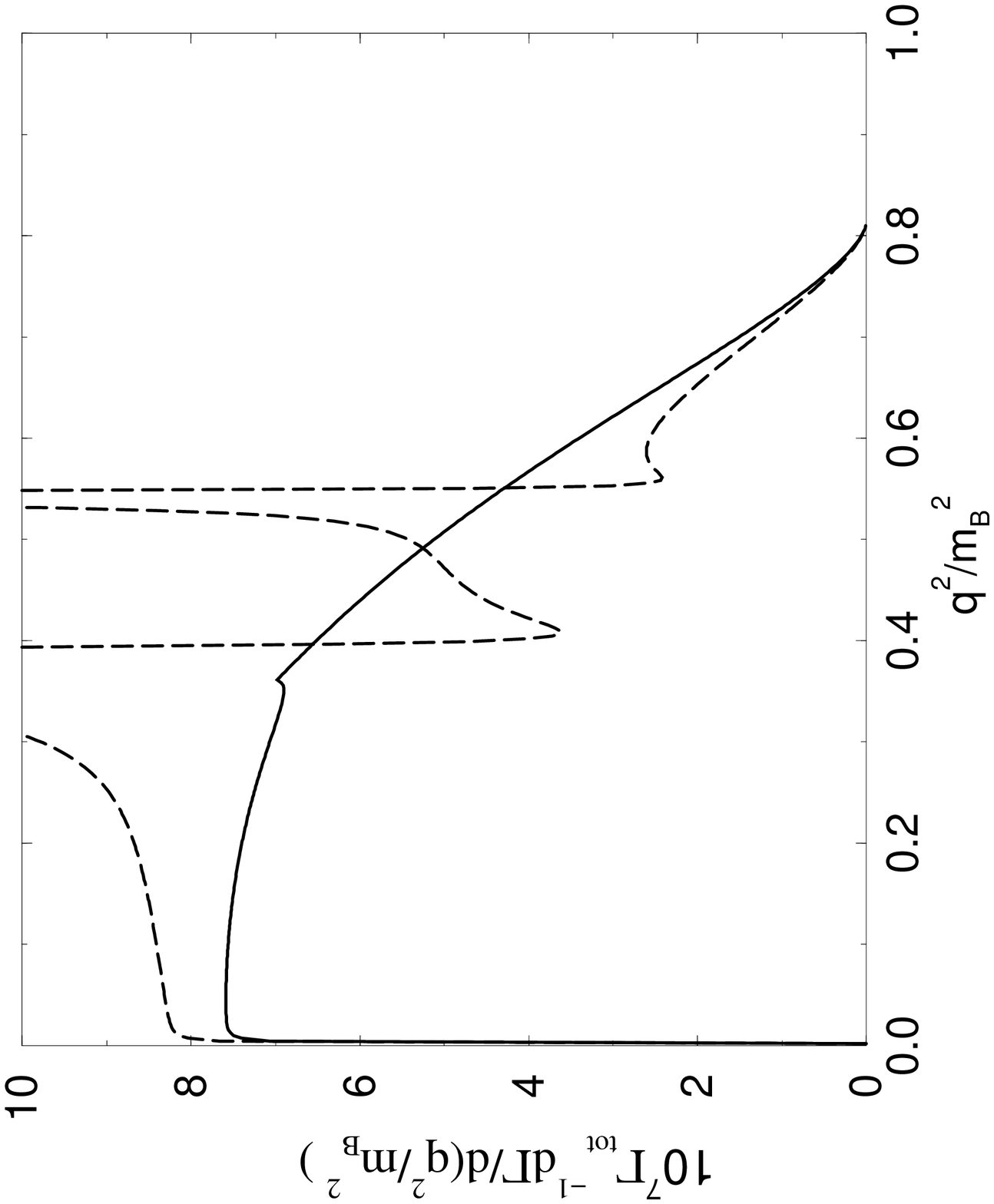}
\end{figure}
\vskip -1.5cm
\hspace{11cm}{\Large a}
\vskip 6.3cm
\begin{figure}[h]
\includegraphics{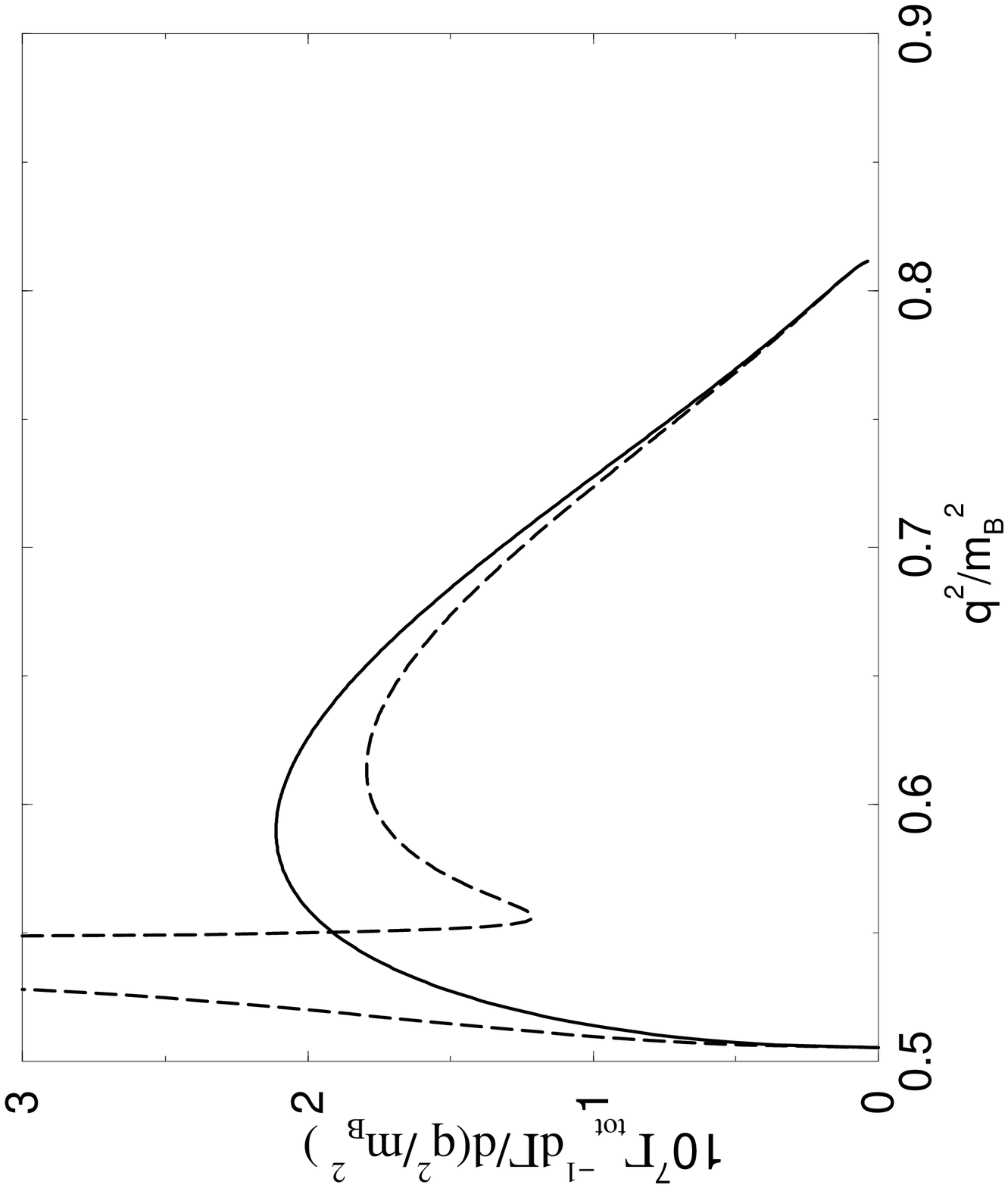}  
\end{figure}  
\vskip 3cm
\hspace{11cm}{\Large b}
\vskip 9.2cm
Fig. 1.
Differential branching ratios for (a) $B\to K\m^+\m^-$ 
and (b) $B\to K\t^+\t^-$ as a function of $\shat=q^2/m_B^2$ 
with $m_t=180\ GeV$. The dashed and solid lines correspond to the results 
with and without the LD contributions from the resonance states, 
respectively. 

\newpage
\begin{figure}[h]
\includegraphics{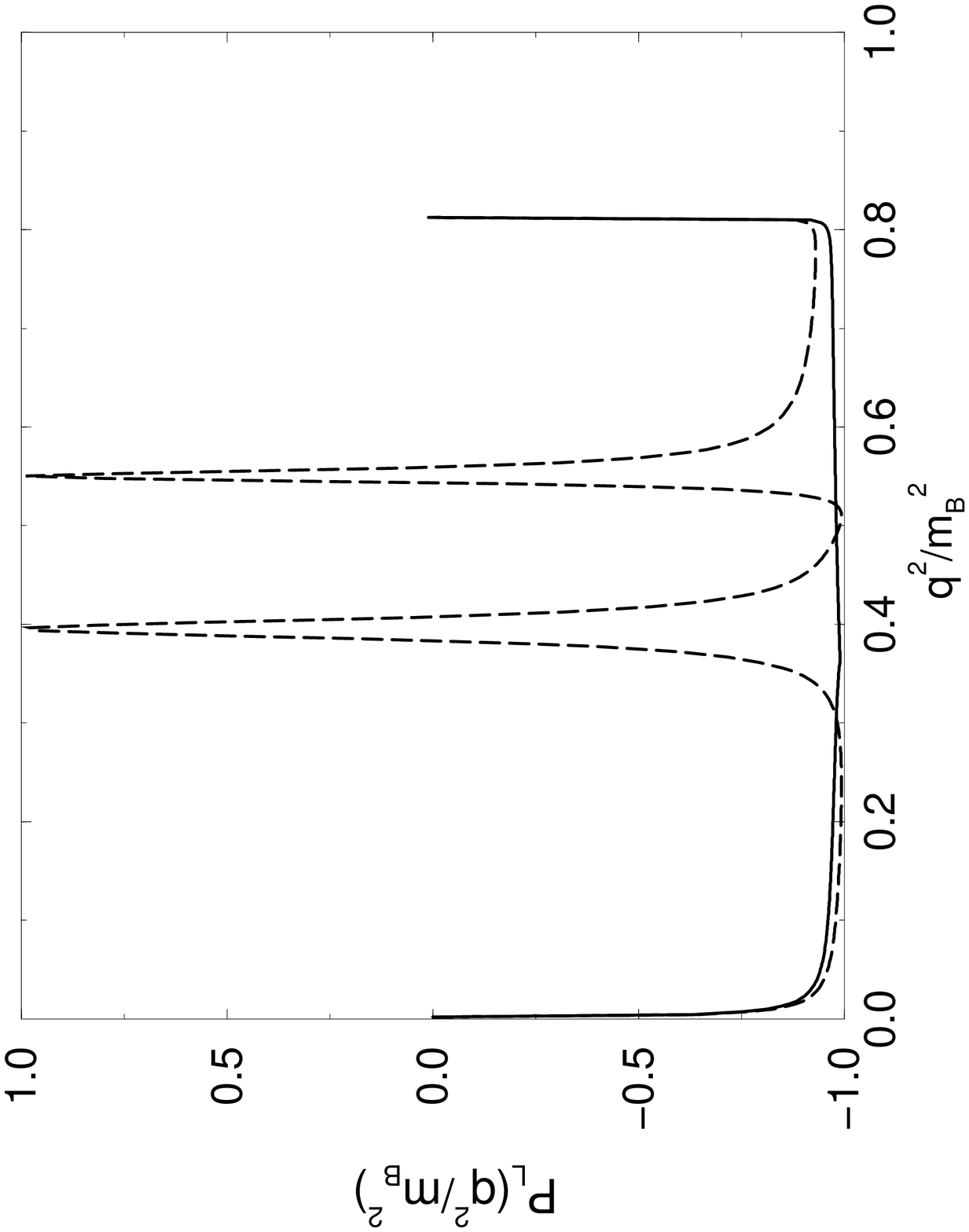}  
\end{figure}  
\vskip -1.5cm
\hspace{11cm}{\Large a}
\vskip 6.3cm 
\begin{figure}[h]
\includegraphics{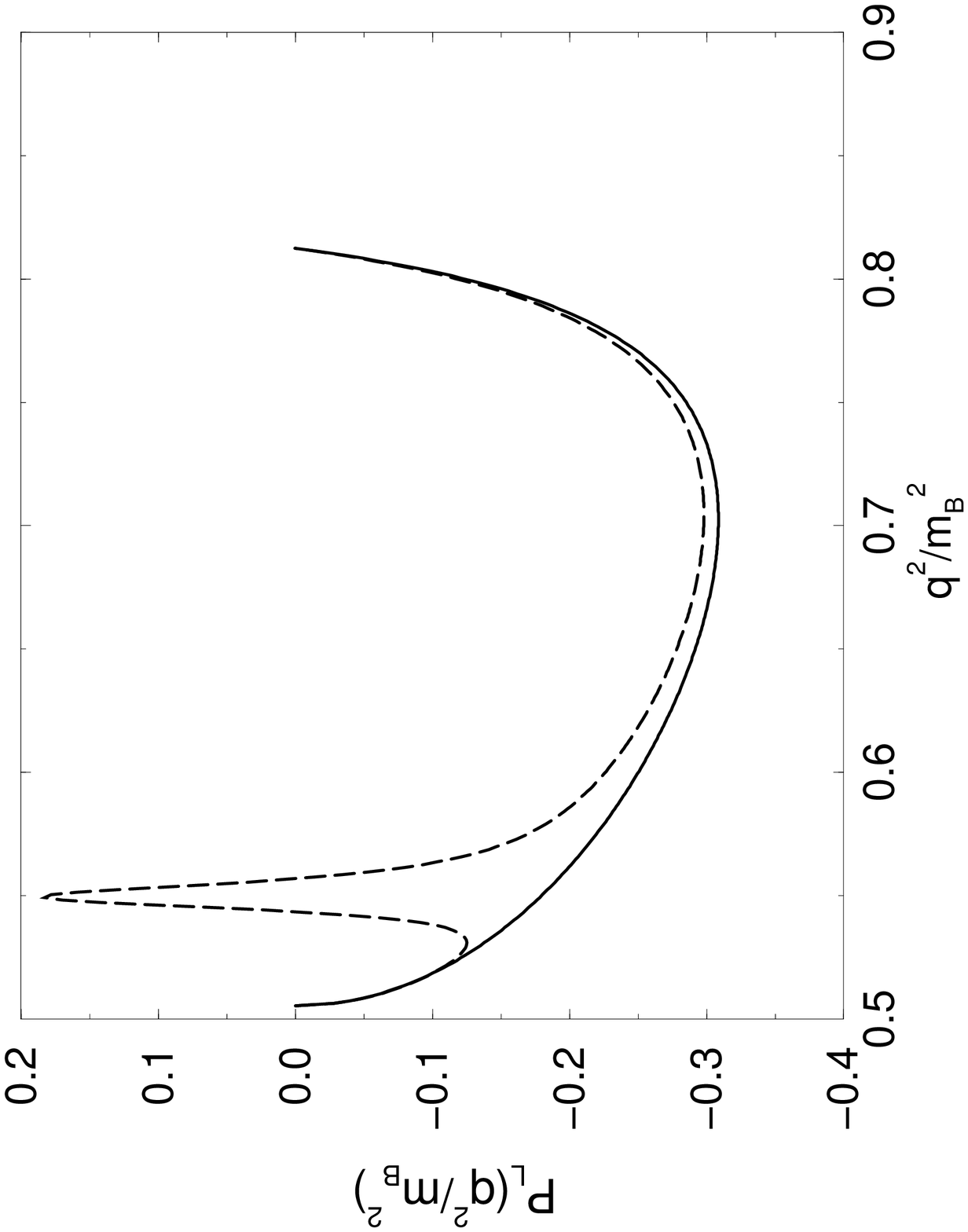}
\end{figure}
\vskip 3cm
\hspace{11cm}{\Large b}
\vskip 9.7cm
Fig. 2. Longitudinal Polarization asymmetries of (a) the muon in
$B\to K\m^+\m^-$ and (b) the tau in $B\to K\t^+\t^-$ as a function 
of $\shat=q^2/m_B^2$ with $m_t=180\ GeV$. 
Legend is the same as in Fig. 1.

\newpage
\begin{figure}[h]
\includegraphics{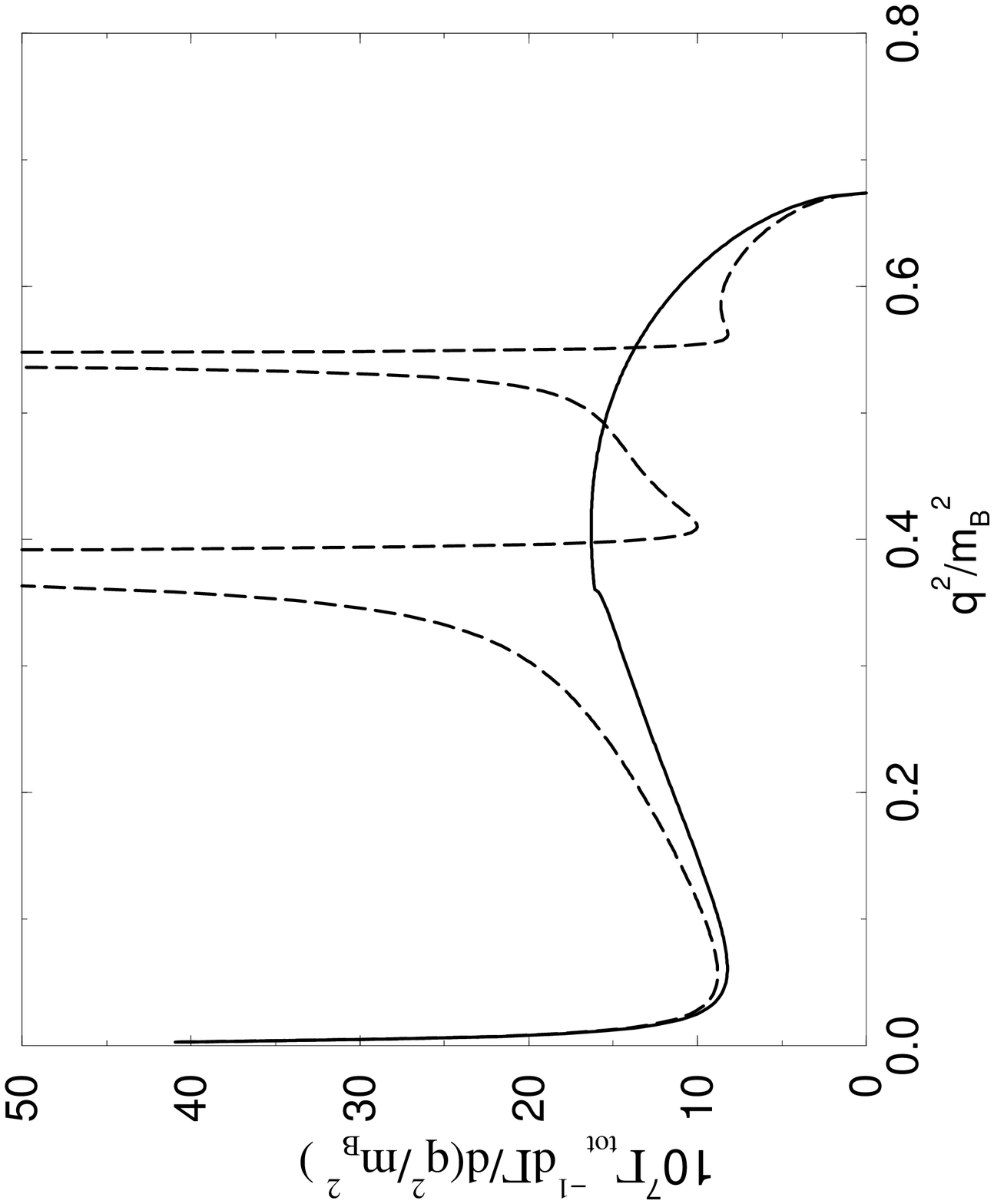}
\end{figure}
\vskip -1.5cm
\hspace{11cm}{\Large a}
\vskip 6.3cm

\begin{figure}[h]
\includegraphics{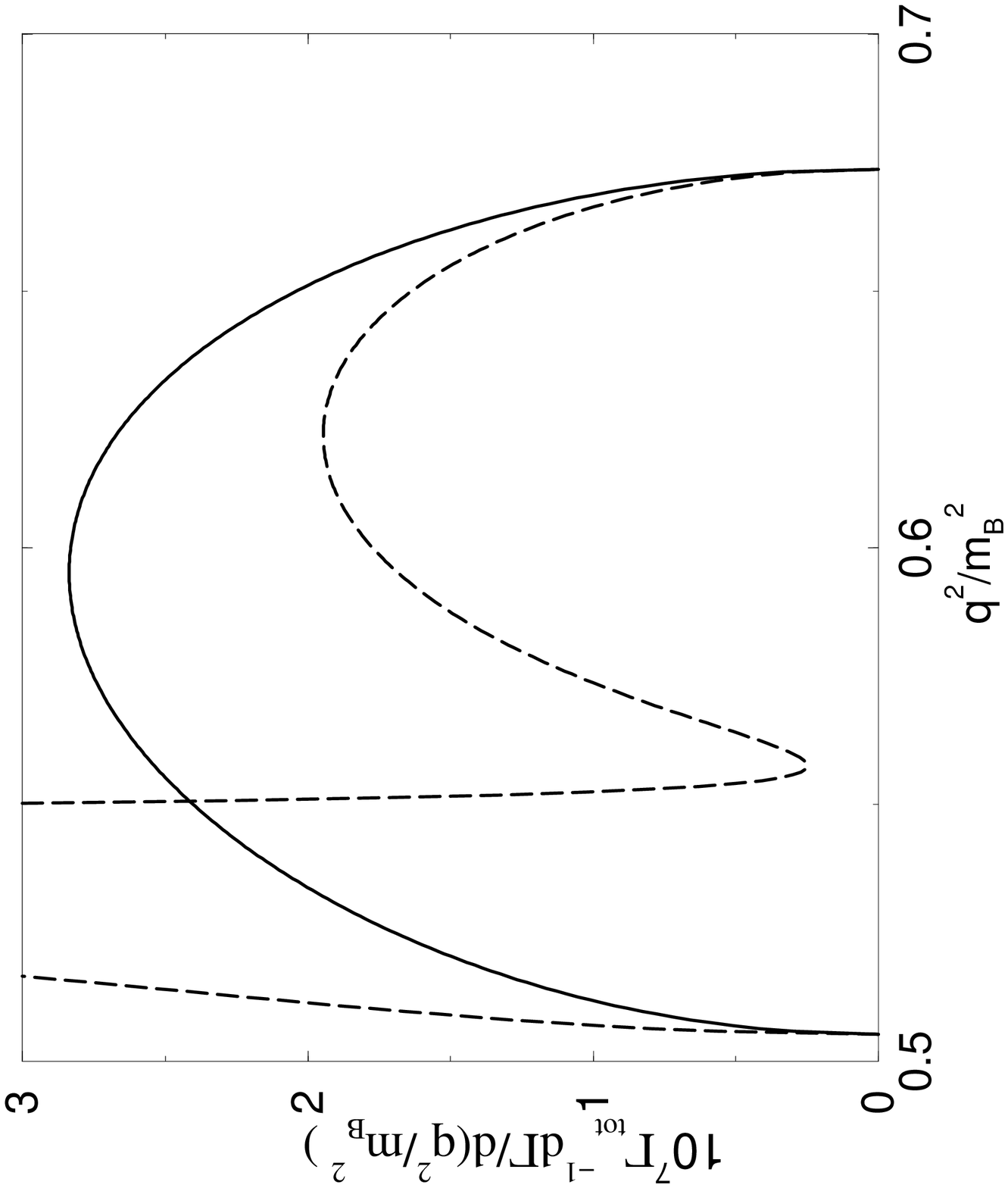}
\end{figure}
\vskip 3cm
\hspace{11cm}{\Large b}
\vskip 9.7cm
Fig. 3.
Same as Fig. 1 but for (a) $B\to K^*\m^+\m^-$ 
and (b) $B\to K^*\t^+\t^-$.

\newpage
\begin{figure}[h]
\includegraphics{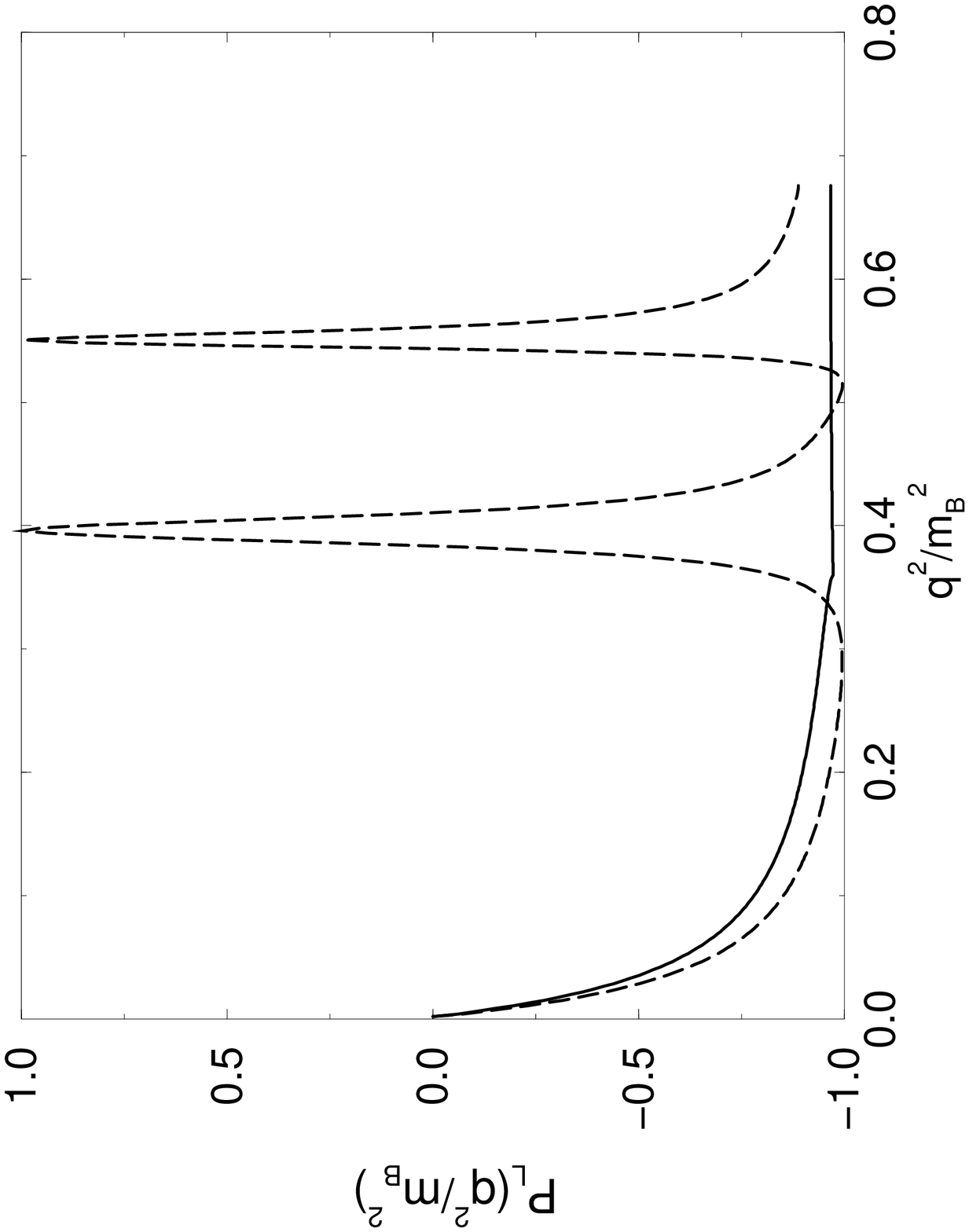}
\end{figure}
\vskip -1.5cm
\hspace{11cm} {\Large a}
\vskip 6.3cm

\begin{figure}[h]
\includegraphics{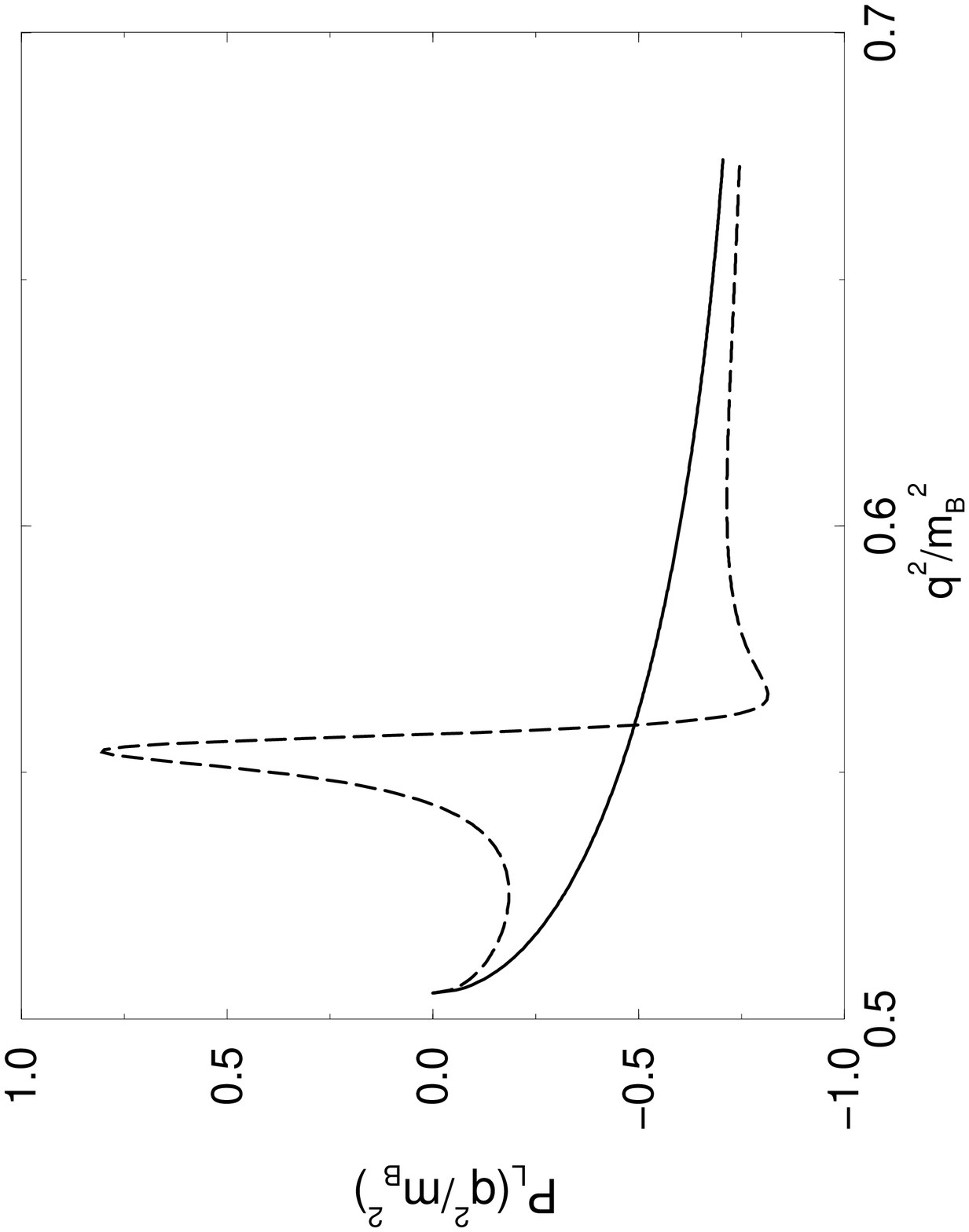}
\end{figure}
\vskip 3cm
\hspace{11cm} {\Large b}
\vskip 9.7cm
Fig. 4.
Same as Fig. 2 but for (a) $B\to K^*\m^+\m^-$ 
and (b) $B\to K^*\t^+\t^-$.

\end{document}